\documentclass[12pt,preprint]{aastex63}

\shorttitle{Improved background of LAXPC}
\shortauthors{Antia et al.}

\begin{document}

\title{Improved background model for the Large Area X-ray Proportional Counter (LAXPC) instrument on-board AstroSat}


\author{}


\author[0000-0001-7549-9684]{H. M. Antia}
\affiliation{UM-DAE Centre of Excellence for Basic Sciences, University of Mumbai, Kalina, Mumbai 400098, India} 
\email{antia@tifr.res.in}
\author{P. C. Agrawal}
\affiliation{Tata Institute of Fundamental Research (retired), Homi Bhabha Road, Mumbai 400005, India}
\author[0000-0002-9418-4001]{Tilak Katoch}
\affiliation{Tata Institute of Fundamental Research, Homi Bhabha Road, Mumbai 400005, India}
\author[0000-0003-0591-9668]{R. K. Manchanda}
\affiliation{Department of Physics and Astronomy, University of Southern Queensland, QLD 4300, Australia}
\author[0000-0003-4437-8796]{Kallol Mukerjee}
\affiliation{Tata Institute of Fundamental Research, Homi Bhabha Road, Mumbai 400005, India}
\author[0000-0002-8016-4077]{Parag Shah}
\affiliation{Tata Institute of Fundamental Research, Homi Bhabha Road, Mumbai 400005, India}




\begin{abstract}
We present an improved background model for the Large Area X-ray
Proportional Counter (LAXPC) detectors on-board AstroSat. Because of
the large collecting area and high pressure, the LAXPC instrument
has a large background count rate, which varies during the orbit. Apart from
the variation with latitude and longitude during the orbit there is a prominent
quasi-diurnal variation which has not been modelled earlier. 
Using over 5 years of background observations, we determined the period
of the quasi-diurnal variation to be 84495 s and using this period, it
is possible to account for the variation and also identify time intervals
where the fit is not good. These lead to a significant improvement in
the background model.
The quasi-diurnal variation can be ascribed to the changes in
charged particle flux in the near Earth orbit.
\end{abstract}



\section{Introduction}

The Large Area X-ray Proportional Counter (LAXPC) instrument aboard the
Indian Astronomy mission AstroSat consists of 3 co-aligned large area proportional counter
units for X-ray timing and spectral studies over an energy range of
3--80 keV \citep{agr06,yad16, agr17}. The detailed calibration of LAXPC
instrument, including the background model was described by \citet{antia17, antia21}.
AstroSat was launched on 28 September 2015 and has completed six years in
orbit.
Currently, only one LAXPC detector, i.e., LAXPC20 is working nominally.
Because of large size of the detector and a gas pressure of two atmospheres,
the background rate is rather high, being about 200 c s$^{-1}$ in LAXPC20.
There is also a strong variation in the background rate during each orbit
with the maximum count rate being achieved near the South Atlantic Anomaly (SAA)
passage. A large part of the orbital variation can be modelled by fitting the
count rate as a function of latitude and longitude of the satellite \citep{antia17, antia21}. Apart from this, there is a quasi-diurnal variation with an
amplitude of about 20 c s$^{-1}$ \citep{antia21} which is not accounted for by
the existing background models. This limits the effectiveness
of background subtraction and hence the sensitivity of the instrument
for faint sources as well as spectral studies at high energies for relatively
bright sources. There is also a long term variation over the period of 6 years
in the background count rate as well as the amplitude of quasi-diurnal trend.

An alternate background model has been implemented by \citet{misra21}
which is applicable only for faint sources and is restricted to the top layer of
the detector. This model is based on the assumption that for faint sources
the counts at high energies are only contributed by the background, which can
be scaled to get the background at low energies. This model is sensitive to
gain shift in the detectors and hence is only applicable to LAXPC20 which has
relatively stable gain. This happens to be the only detector which has been working
nominally after April 2018. This model can account for
the quasi-diurnal trend to some extent, but its applicability is limited
to faint sources and the top layer of the detector. Further, the calibration
of this model has not been updated and hence it doesn't work so well on
recent observations  and in fact the diurnal variation is not removed
\citep[Figure 15 of][]{antia21}.

The main difficulty in accounting for the quasi-diurnal trend was that its
exact period was not known. During the initial years the amplitude of
this signal was also smaller. With increasing amplitude of the trend it
became more important to understand this trend. Using more than 5 years of
background observations and satellite orbital parameters, we have now
identified the period of quasi-diurnal variation to be 84495 s, which allows
us to remove the trend effectively and improve the background model.

The rest of the paper is organized as follows: Section 2 gives a summary of
various periodicities in AstroSat orbit so that the observed periodicity
in the background count rate can be identified with some of these.
Section 3 describes the determination of periodicity in background count rate
using over five years of background observation. Section 4 describes the
application of this to background model and the resulting improvement.
Section 5 gives the summary of the results.

\section{The AstroSat orbit}

AstroSat has been placed in a nearly circular orbit with eccentricity of
$\sim 0.001$, altitude of about 640 km and a inclination of about $6^\circ$
to the equator. The altitude has been decreasing slowly with time, resulting
in a slow decrease in the orbital period. During April 2021, the sidereal
orbital period was 5844.38 s, while the period as measured by latitude
variation was 5836.57 s, that from longitude was 6269.66 s and the altitude
had a period of 5851.71 s. The frequency difference between the latitude and
longitude gives a period of 84494 s. This differs from the sidereal rotation
period of the Earth, 86164 s because of precession of the orbit.

\begin{figure}
\epsscale{.95}
	\plottwo{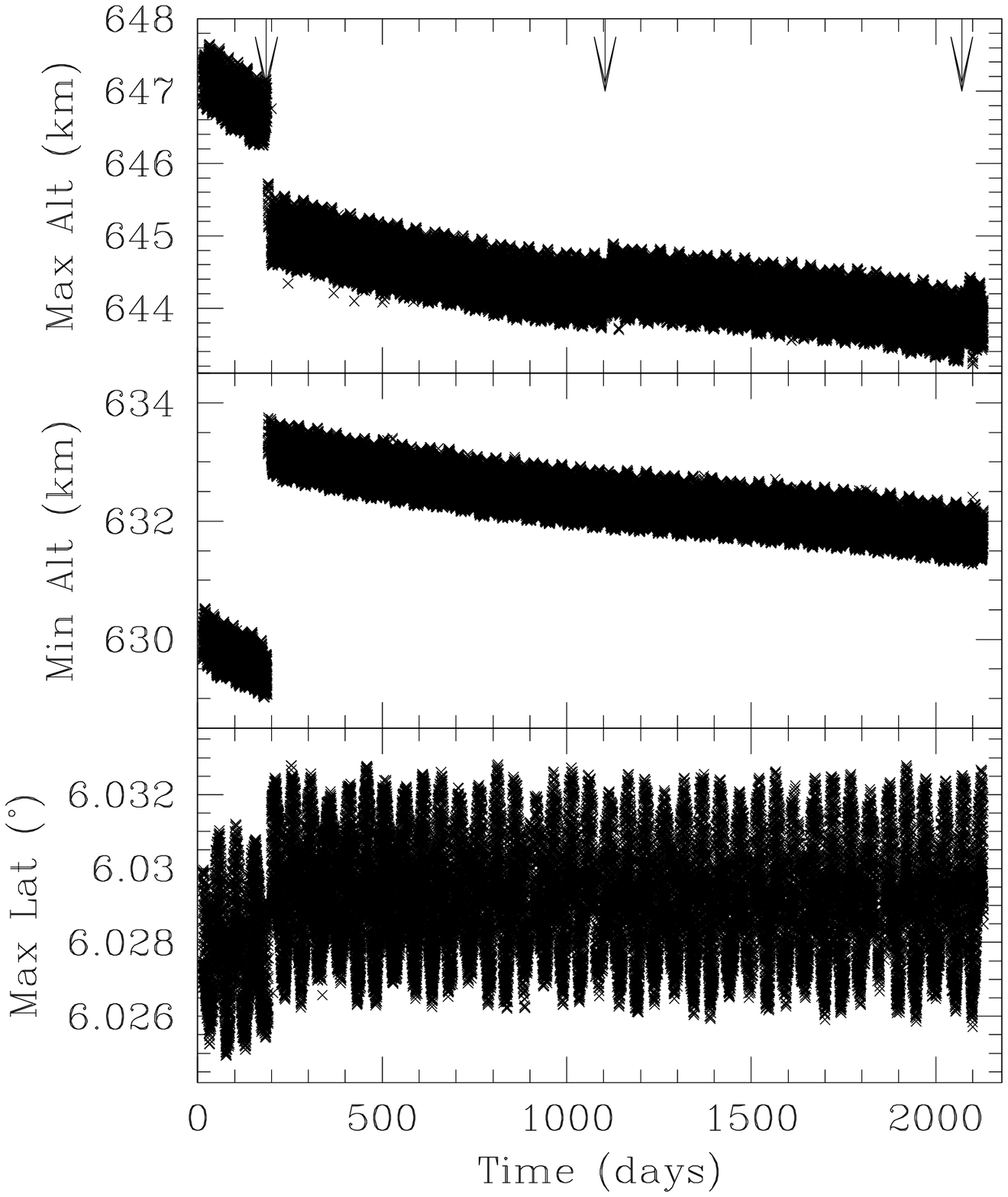}{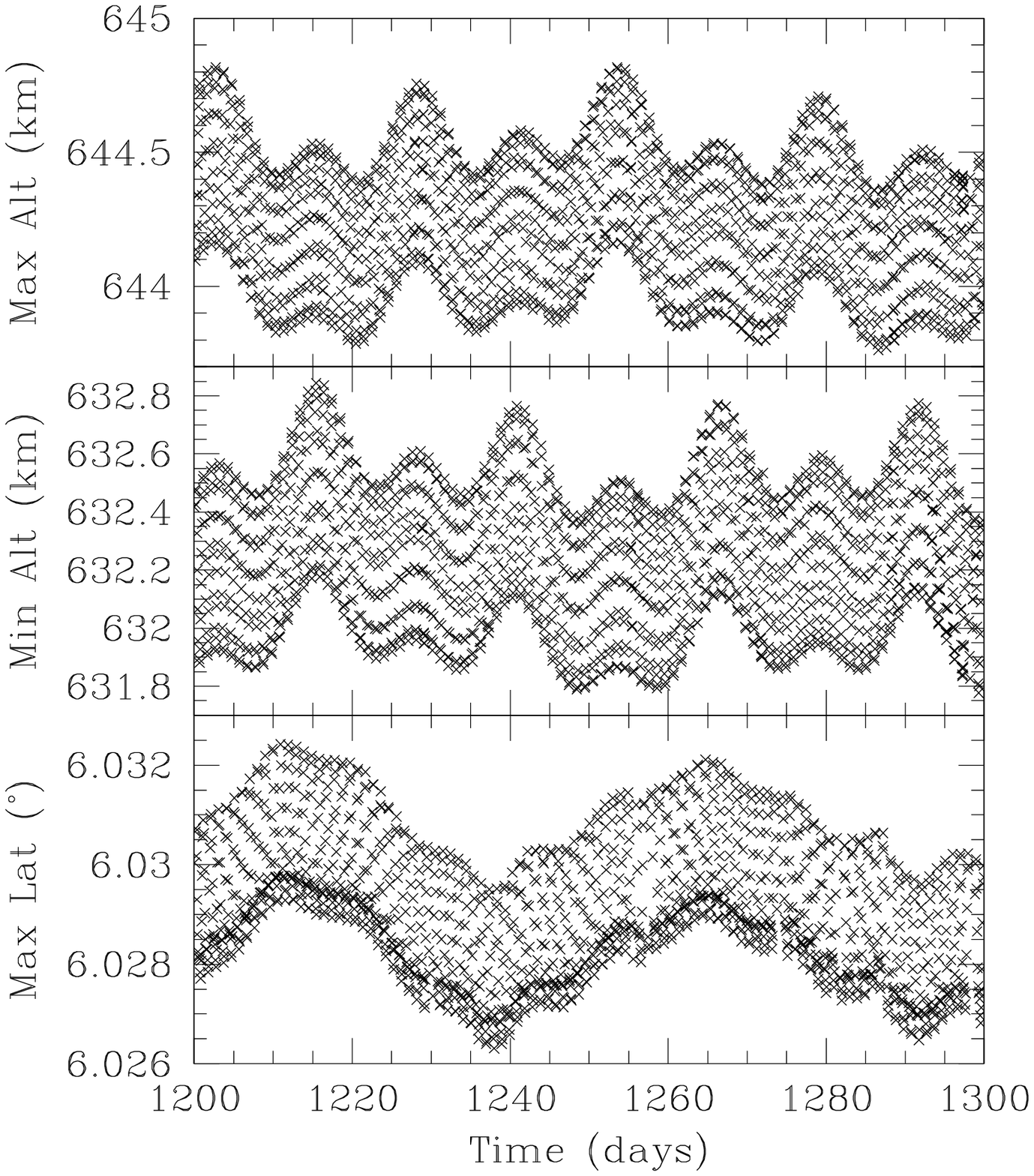}
\caption{The maximum and minimum altitude during each orbit and the maximum
latitude during each orbit, is shown in the left panel as a function of days since the
launch of AstroSat. The arrows in the top panel mark the times of
AstroSat orbit maneuvers. The right panel shows the same quantities over
a limited time to show the variations on scale of days.}
\label{fig:alt}
\end{figure}

Since the eccentricity of the orbit is very small the rate of precession
of the nodes or the plane of the orbit can be estimated by considering the
perturbation due to gravitational quadrupole moment, $J_2$ ($=0.00108$ for the
Earth) to get the precession rate
\begin{equation}
\frac{3\pi R^2}{T a^2}J_2\approx 1.44\times 10^{-6}\;\mbox{rad}\;\mbox{s}^{-1},
\end{equation}
where, $R$ is the radius of the Earth, $a$ is the semi-major axis of the
satellite orbit and $T$ is the orbital period of the satellite. This gives
the precession period of about $4.36\times 10^6$ s. For nearly circular orbit this
is also the period of precession of the orbital plane. Since the maximum
latitude during the orbit is changing with time the orbital plane should be
precessing.

To determine the actual period of precession of AstroSat orbit we examined
the orbital data for over five years and determined the minimum and maximum of
altitude and latitude during each orbit and the results are shown in Figure~\ref{fig:alt}. The minimum in latitude is not shown as it is almost same in magnitude
to the maximum. The right panel of the figure shows the blowup of the same,
which clearly shows variation on quasi-diurnal period as well as variation
on time scale of order of 50 days. For the maximum latitude these periods
averaged over the last five years are 84496 s and 4363843 s. The latter 
should be the precession period. If the corresponding frequency is added
to the Earth's rotation frequency we get a period of 84496 s, which is
close to the frequency difference
between the latitude and longitude periods. This is also close to the period
of quasi-diurnal variation found in the LAXPC background. The three arrows
in the figure mark the times when some orbital maneuvers were carried out.
These maneuvers can change the orbital period, but it was found that the
change in the period of quasi-diurnal variation due to the last two
maneuvers is less than 1~s and hence is not considered. We have not
considered the period before April 4, 2016, when the first maneuver was
performed as that would have changed the period significantly and we do not
have enough data before that to determine the period accurately. There was
also only one long background observation before this time. It is difficult
to model the background during this initial period as there were some
adjustments made to various parameters in the detector.

While analyzing the orbital data it was discovered that during four
periods of about one day  each, there is some anomaly in the orbital
data. Since the LAXPC analysis software relies on the satellite position to
determine the time of Earth occultation and SAA passage these data may
not be processed correctly. The affected observations are listed in
Table~1, which also gives the time interval that is affected in MJD.
The data from this period should be rejected during analysis.
The version v3.4 onward of LaxpcSoft\footnote{https://www.tifr.res.in/\~{}astrosat\_laxpc} (backshiftv3.f) would issue a warning if these data are
processed.

\begin{deluxetable}{llc}
\tablecaption{LAXPC observations affected by anomaly in orbital data}
\tablewidth{0pt}
	\tablehead{\colhead{Observation ID}&\colhead{Source}&\colhead{Affected Period (MJD)}}
\startdata
20180529\_G08\_025T01\_9000002130 & Cyg X-3 & 58268.25 -- 58269.25  \\
20200123\_A07\_138T09\_9000003458 & J164754.90+443345  & 58872.00 -- 58873.30 \\
	20200910\_A09\_079T01\_9000003864 & CAL 83 & 59103.90 -- 59105.00 \\
	20210120\_A10\_123T08\_9000004126 & BCD T8 & 59234.95 -- 59235.90 \\
	20210121\_A10\_093T01\_9000004128 & 2MASS J05215658+4359220 & 59234.95 -- 59235.90   \\
\enddata
\label{tab:ei}
\end{deluxetable}

With the precession period obtained above, all the periodicities in AstroSat orbit
can be explained. If $f_A=1/5844.38$ Hz is the sidereal frequency of AstroSat
orbit, $f_E=1/86164$ Hz is the Earth's rotation frequency and
$f_P=1/4363843$ Hz is the precession frequency of AstroSat orbit, the
frequencies, $f_\mathrm{lat},f_\mathrm{lon},f_\mathrm{alt}$ of the orbital
period in longitude, latitude and altitude, respectively, can be determined
as follows:
\begin{eqnarray}
	f_\mathrm{lon}&=& f_A-f_E= {1\over 6269.64}\; \mathrm{Hz},\\
	f_\mathrm{lat}&=& f_A+f_P= {1\over 5836.56}\; \mathrm{Hz},\\
	f_\mathrm{alt}&=& f_A-f_P= {1\over 5852.22}\; \mathrm{Hz},
\end{eqnarray}
These values are all close to the observed periods mentioned earlier.
Further,
\begin{equation}
	f_\mathrm{lat}-f_\mathrm{lon}=f_P+f_E={1\over 84496}\;\mathrm{Hz}.
\end{equation}
This is close to the actual frequency difference as noted earlier.
As shown in the next section this is also close to the period of quasi-diurnal
variation in the LAXPC background.

\section{The period of quasi-diurnal variation in the background}

To determine the period of quasi-diurnal variation in the background
we use all background observations after April 4, 2016 covering
a stare time of approximately a day or more. The light curves of all
these observations with a time-bin of 10 s were combined into one
time series. Only the Good Time Intervals (GTI) were considered, which exclude the passage
through the South Atlantic Anomaly (SAA) and when the satellite is pointing
to the Earth as the target location is occulted by the Earth. The dominant
variation in the time-series is due to the orbital period of longitude,
which gives the maximum count rate when the satellite is close to the
SAA passage. Apart from this the quasi-diurnal oscillations are also seen
\citep{antia21}. To fit both these periodic variations we fit the combined
time series, $c(t)$, to the function
\begin{equation}
	c(t)=a_0+\sum_{j=1}^N\left(a_j\cos(2\pi jf_1t)+b_j\sin(2\pi jf_1t)
	+d_j\cos(2\pi jf_2t)+e_j\sin(2\pi jf_2t)\right),
\end{equation}
where, $f_1,f_2$ are respectively, the frequencies of the orbital and quasi-diurnal
variations and $N$ is the number of harmonics used in the fits.
The parameters $a_j,b_j,d_j,e_j$ and frequencies $f_1,f_2$ are fitted to
minimize the resulting $\chi^2$ deviation. $N=5$ was used in the fits as
that was found to be sufficient.

\begin{figure}
\epsscale{.95}
\plotone{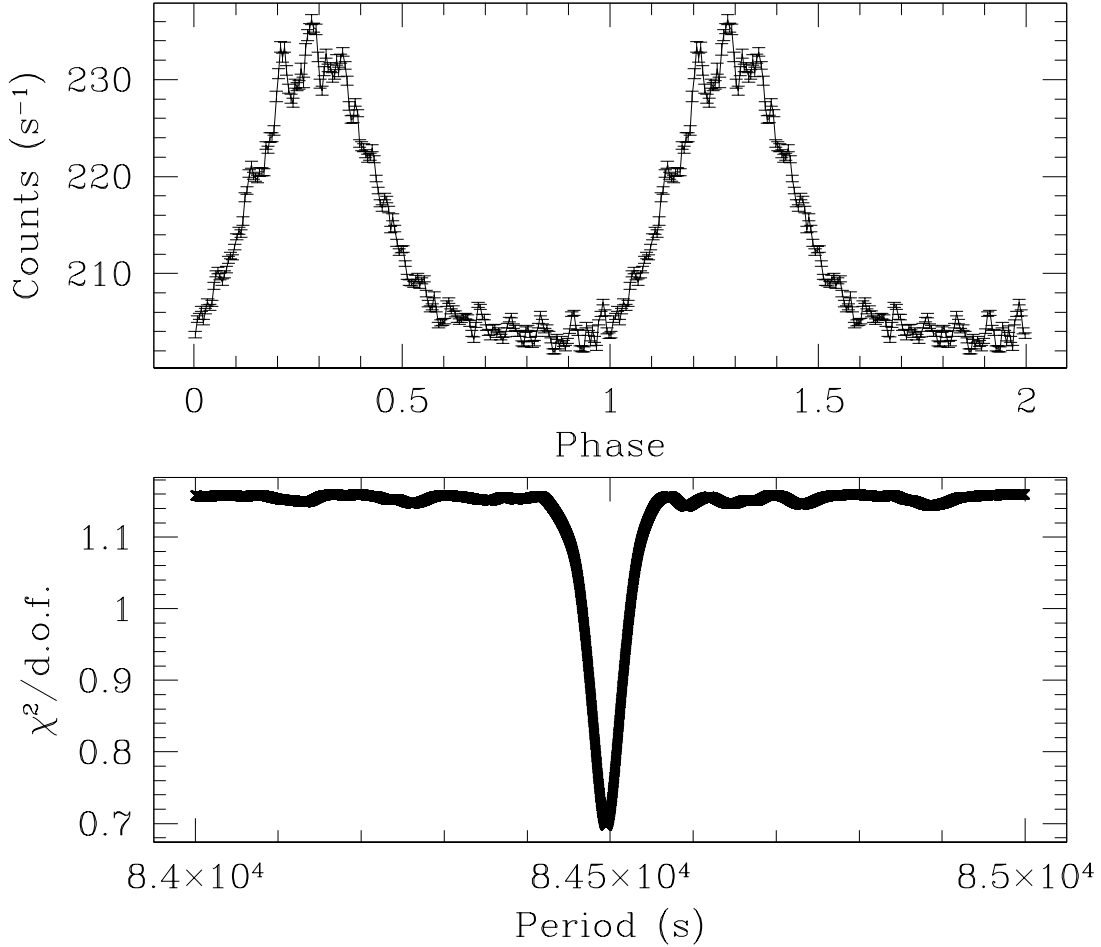}
\caption{The least square fit to background light curves, showing the
reduced $\chi^2$ as a function of the quasi-diurnal period in the bottom
panel and the resulting profile as a function of phase in the top panel.}
\label{fig:diur}

\end{figure}
\begin{figure}
\epsscale{.95}
\plotone{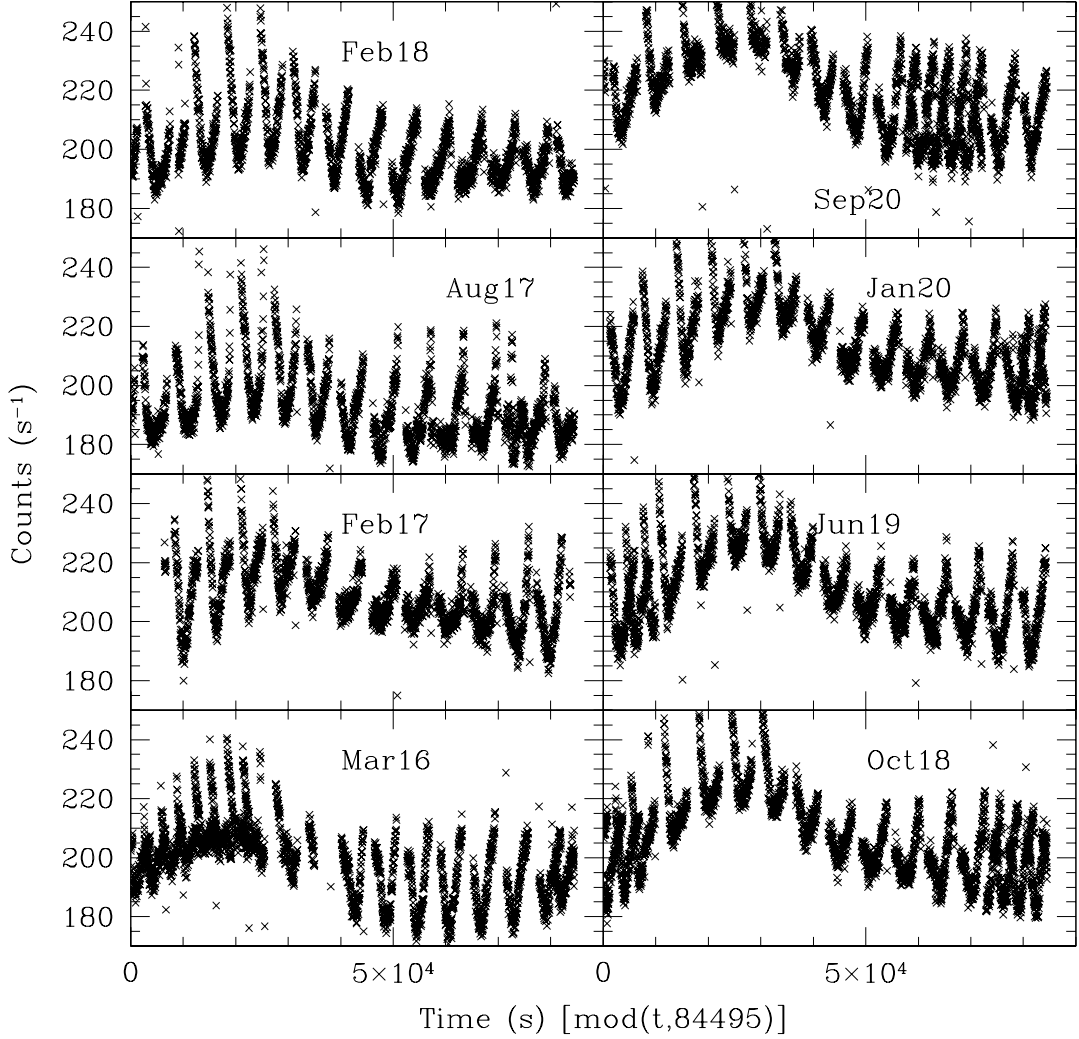}
\caption{The light curves for a few selected background observations in
	LAXPC20 with a time-bin of 32~s are shown as a function of $\mathrm{mod}(t,84495\;{\rm s})$.
	Since the background
	observations cover a period longer than 84495 s, there is some
	time period where multiple values are available.}
\label{fig:back}
\end{figure}

The resulting least squares fit gives the two periods as 6270.40 s and
84495 s. The statistical errors in the fits were small and hence are not shown.
The value of 6270.40 s is in agreement with the period in
longitude. Likewise, the second period of 84495 s agrees with the period of quasi-diurnal variation in the maximum of
latitude obtained in the previous section.
Figure~\ref{fig:diur} shows the $\chi^2$ value as a function of 
the quasi-diurnal period and the resulting profile of diurnal variation.
It is clear that there is a very clear minimum in $\chi^2$ giving a reliable
estimate of the period of quasi-diurnal variation. The profile obtained
matches with the observed profile as shown in Fig.~\ref{fig:back} which
shows the light curves for a few selected background observations.
The time axis in this figure is $\mathrm{mod}(t,84495)$ and it can be seen that
in all cases spanning about 5 years the maximum occurs around the same phase,
thus confirming that the variation has maintained the phase over the
entire duration of AstroSat observations. This can be compared with
Figure~12 of \citet{antia21} which shows the light curves for the same
background observations with time measured from the beginning of observation.
That figure naturally shows peak in quasi-diurnal variations at different times.

From the discussion in the previous section it is clear that this is essentially
the diurnal variation due to the rotation of the Earth. The period differs
from the rotation period of the Earth because of precession of the
AstroSat orbit.

\section{Improved background model}

The observed background light curve for each background observation
is fitted to a function of latitude
and longitude to account for the variation during the orbit. To account
for the quasi-diurnal variation an additional periodic function of time
is added in the improved model to get
\begin{equation}
	c(t)=\sum_{ij}g_{ij}\phi_i(\theta)\psi_j(\phi)+\sum_{j=1}^N
	\left(d_jcos(2\pi f_2j t)+e_j\sin(2\pi f_2jt)\right),
	\label{eq:backfit}
\end{equation}
where $\theta,\phi$ are the latitude and longitude of satellite position
at time $t$, $f_2$ is the frequency of the quasi-diurnal variation determined
in the previous section and $\phi_i(\theta)$ and $\psi_j(\phi)$ are the
cubic B-spline basis functions over a suitably chosen knots in latitude and
longitude, respectively. The coefficient $g_{ij}, d_j, e_j$ are determined by
a regularized least squares fit \citep[e.g.,][]{ant12} to the observed light curve.
A time bin of 32 s was used in the light curve.
The number of harmonics, $N=5$ is used, which is the same as that used for
fitting the period of quasi-diurnal period in the previous section.
In order
to cover the entire range of latitude and longitude, all background
observations extend over a period of more than one day.
The Eq.~\ref{eq:backfit} defines the background model which is implemented
in the LAXPC analysis software (laxpcl1.f) which is available from the LAXPC
website\footnote{https://www.tifr.res.in/\~{}astrosat\_laxpc}. The
LaxpcSoft v3.4.2 onward implement the revised background model.

\begin{figure}
\epsscale{.95}
\plottwo{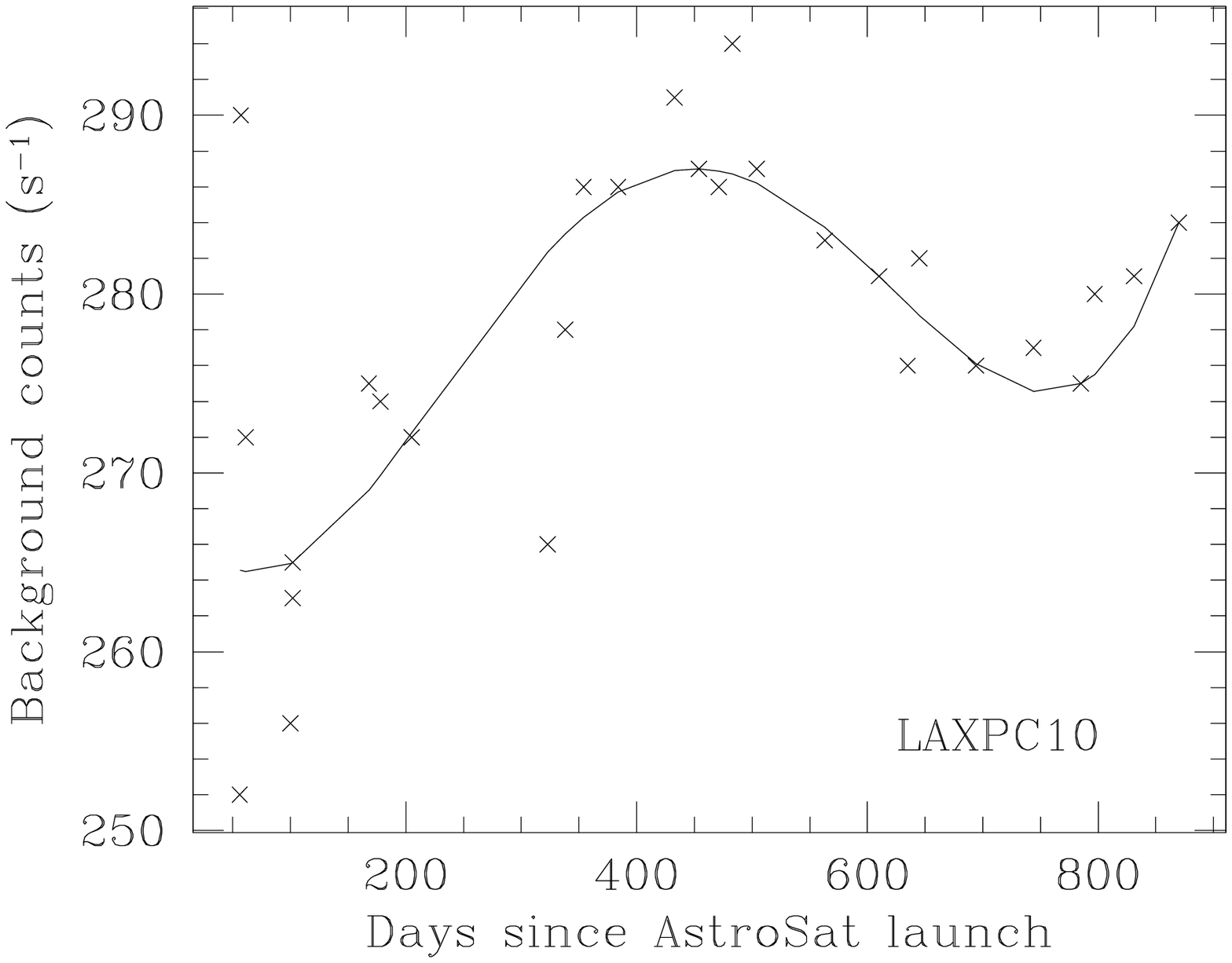}{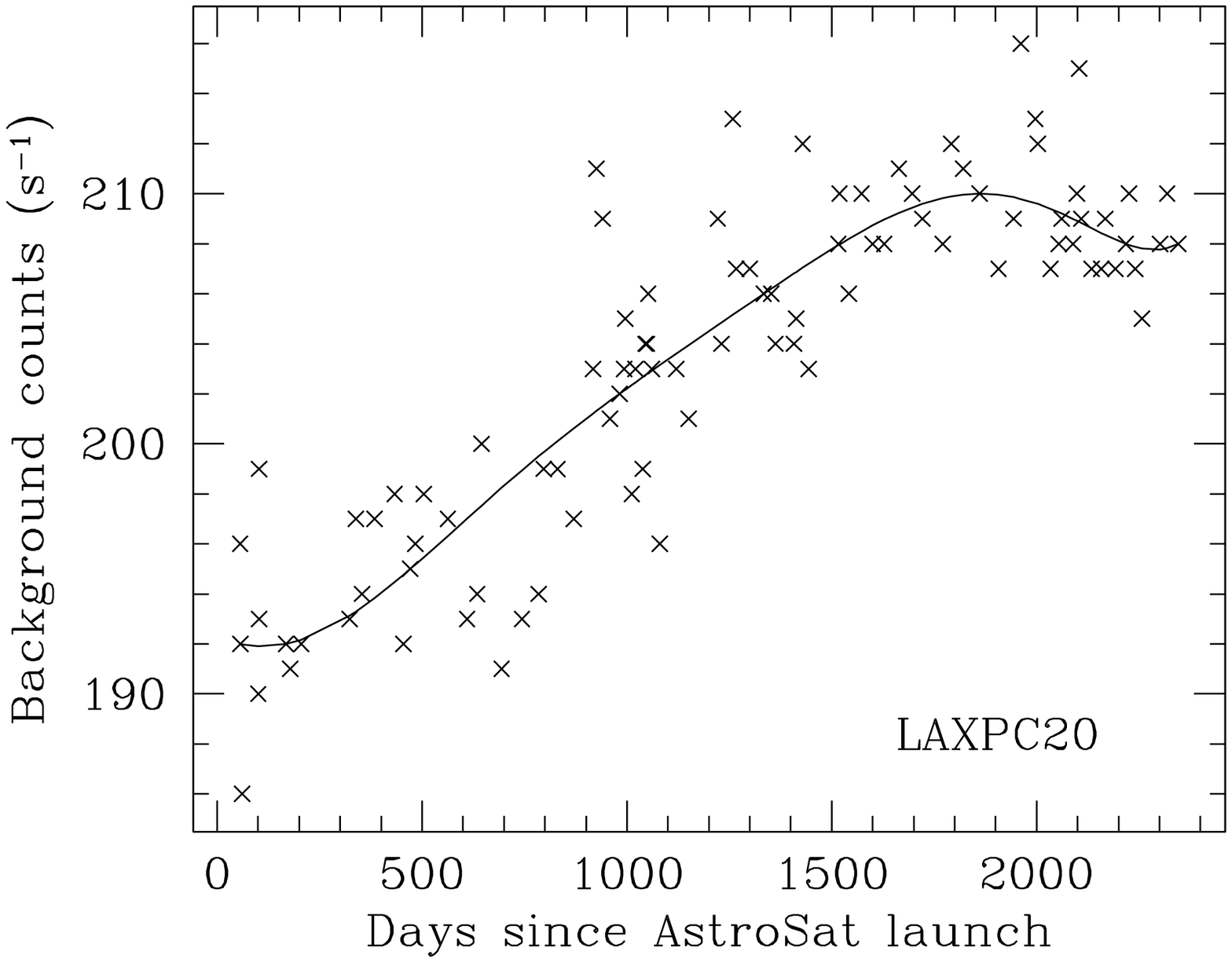}
\caption{The long term variation in the background count rate after
correcting for gain shift and quasi-diurnal variations in LAXPC10 (left panel) and LAXPC20 (right panel).}
\label{fig:backlong}
\end{figure}

\begin{figure}
\epsscale{.98}
\epsfclipon
\plottwo{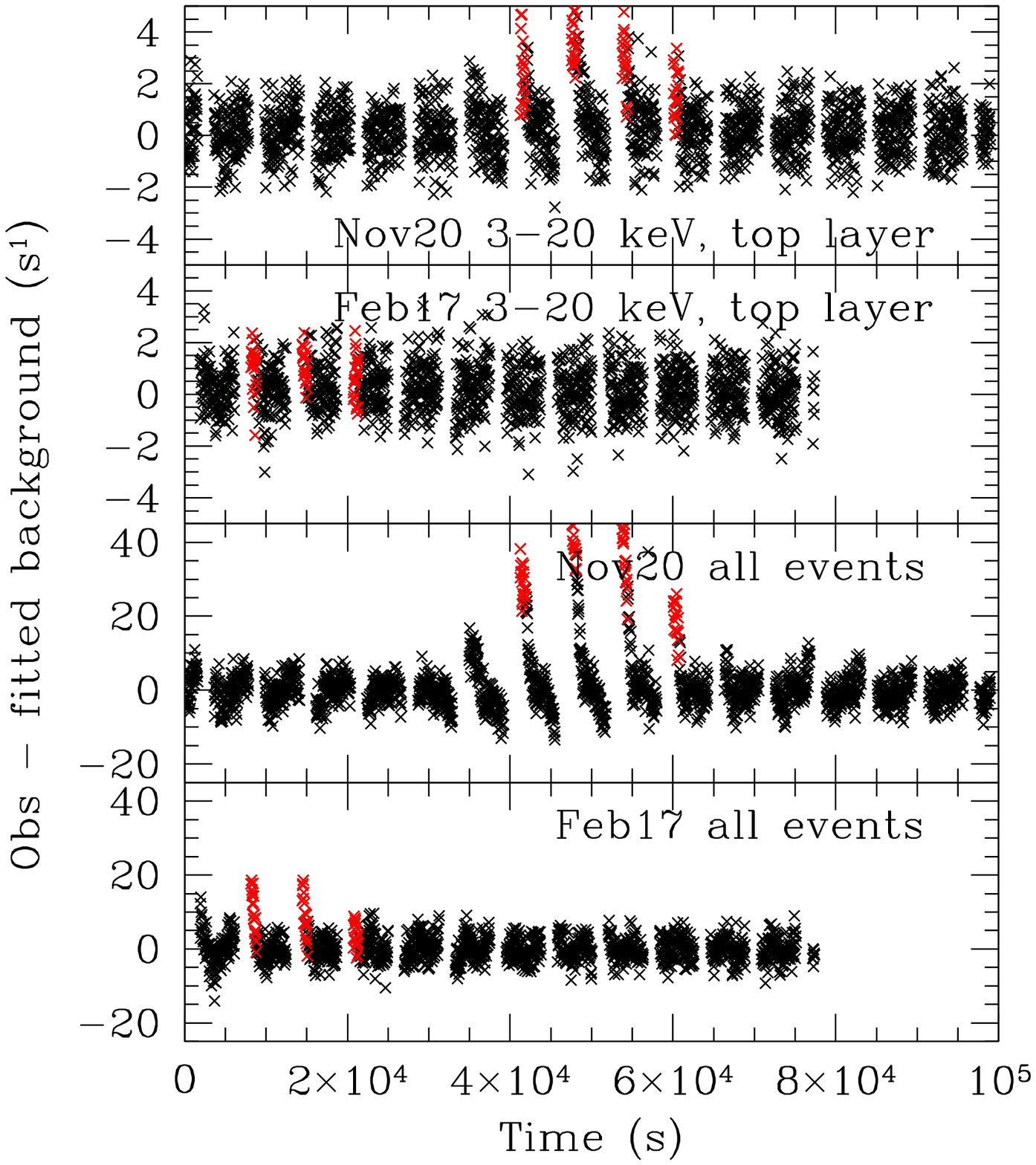}{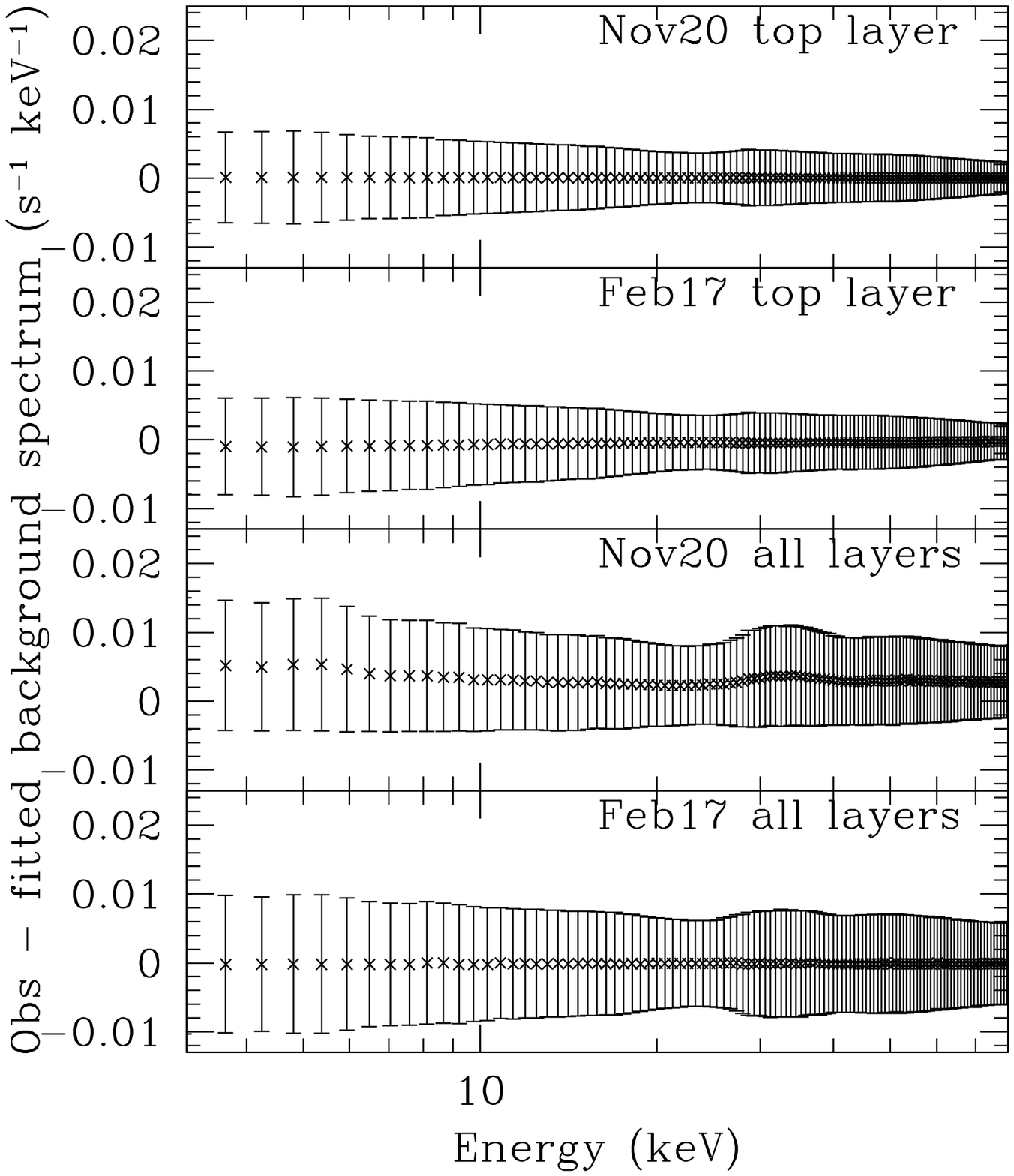}
\caption{The residuals in the fit to background of LAXPC20 for the two
background observations obtained using the improved background model.
The left panel shows the residuals in the light curve with a time-bin of 32 s.
The red points mark the time interval that is now removed from the GTI.
The right panel shows the residuals in the energy spectrum.}
\label{fig:backfit}
\end{figure}

Even after including the quasi-diurnal variation, the background model
defined above does not fit the observed variation (Figure~\ref{fig:backfit})
over the entire region.
There is a significant deviation in the few orbits near the maximum in
quasi-diurnal variation, just after the satellite exits the SAA. This region also
shows variation in the spectrum \citep{antia21}. Thus in the revised
software a period of 10 min after the exit from SAA in orbits close to the
maximum in quasi-diurnal variation is removed from the Good Time Interval (GTI).
These orbits are determined by the time, $t$ (in sec), of exit from SAA, which
satisfies, $10000\le{\rm mod}(t,84495)\le35000$. The inclusion of quasi-diurnal
variation and the improved definition of GTI leads to a significant
improvement in background model as seen in the following discussion.

Apart from the orbital and quasi-diurnal variation, the LAXPC background
also shows some long term variation as shown in Figure~\ref{fig:backlong}.
This figure shows the average count rate during the background observations
corrected for gain shift and quasi-diurnal variation.
It can be seen that the background counts in LAXPC20 have increased by
about 20\% over the last 6 years, but the gradient has reduced since
2019 and the counts are relatively stable during 2020--21.
Figure~\ref{fig:backfit} shows
the residuals in the background fit to the light curve and the spectrum
for two representative background observations. This figure can be
compared with Figure~14 of \citet{antia21} which shows the results for the
same background observations using old background model. It can be seen
that there is some improvement in the fit to both the light-curve and the
spectrum. It may be noted that the $y$-axis range in the spectrum fit is
reduced as compared to that in the old figure due to significant improvement
in the fit. Since the typical count rate in LAXPC spectrum is 2 c s$^{-1}$
keV$^{-1}$, the difference is less than 1\% in all cases.
For the February 2017 background observation it appears that there was no need to restrict
the GTI. This is a special case, while for almost all background observations
it is needed to remove some part of GTI. The reason for this is not clear,
but out of about 60 background observations only in 3 cases the GTI adjustment may
not be needed. For the sake of uniformity we have applied this correction to
all observations after 4 April 2016. For earlier observations the GTI correction
is not applied. There is only one long background observation during March 2016
in this earlier period.

As described by \citet{jah06} the detector background is expected to be contributed by
(i) the cosmic X-ray background, (ii) charged particle flux in the local
environment (iii) induced radioactivity in the spacecraft. Simulation of
the detector background \citep{antia17} appears to suggest that the first component is
dominant as far as total count rate is concerned. However, the temporal
variations are predominantly caused by the other two components. The
secular long term variation are likely to be from the induced radioactivity,
while the shorter term variations on time-scale of a day or shorter may be
from the charged particle flux in the satellite environment.
This component is likely to be correlated with the geographic position
of the satellite relative to the Earth. As a result, we can expect this
variation to show diurnal variation with the period estimated above.
In fact, the region with high background count rate in LAXPC is correlated to region where the
Charged Particle Monitor (CPM) on board AstroSat \citep{rao17} finds high count rate.
The Proportional Counter Array (PCA) on board the Rossi X-ray Timing
Explorer (RXTE) also appears to show similar variation \citep[Figure 25 of][]{jah06}
which has been attributed to the correlation with apogee precession
\citep{jah06}. Similarly, the Large Area Counter (LAC) on board Ginga also
shows similar variations \citep{hay89}.
This appears to be similar to what we find in the LAXPC
detector.
The orbital parameters of Ginga/LAC, RXTE/PCA and AstroSat/LAXPC are quite
different. Ginga and RXTE had higher orbital inclination of  $31^\circ$ and
$23^\circ$ \citep{jah96}, respectively, while AstroSat is in near equatorial
orbit with an inclination of $6^\circ$. The Ginga orbit also had a higher
eccentricity with altitude ranging from 517 km to 708 km. Similarly, RXTE
altitude ranged from 580 km just after launch to 490 km at the end of operation.
While AstroSat altitude is in a narrow range of 630 km to 650 km (Figure 1).
AstroSat passes through the northern end of SAA where the
charged particle flux is order of magnitude or more lower than that for latitudes of
$15^\circ$ to $30^\circ$, traversed by Ginga and RXTE.
As a result, we would expect the induced radioactivity in AstroSat to be
much lower than that in Ginga or RXTE. The short term variations in
the background rates of LAC, PCA and LAXPC are therefore, expected to show
some differences, but the basic features of the background are the same.
The short term variations in LAC and PCA have been modelled 
using induced radioactivity. For LAXPC we have been able to model most
of the variation by treating the count rate as a function of latitude and
longitude, supplemented by a diurnal variation. The remaining part in a
few orbits is discarded from the GTI as that shows a rather large deviation,
which has significant variation on long time-scales.

\begin{figure}
\epsscale{.70}
\plotone{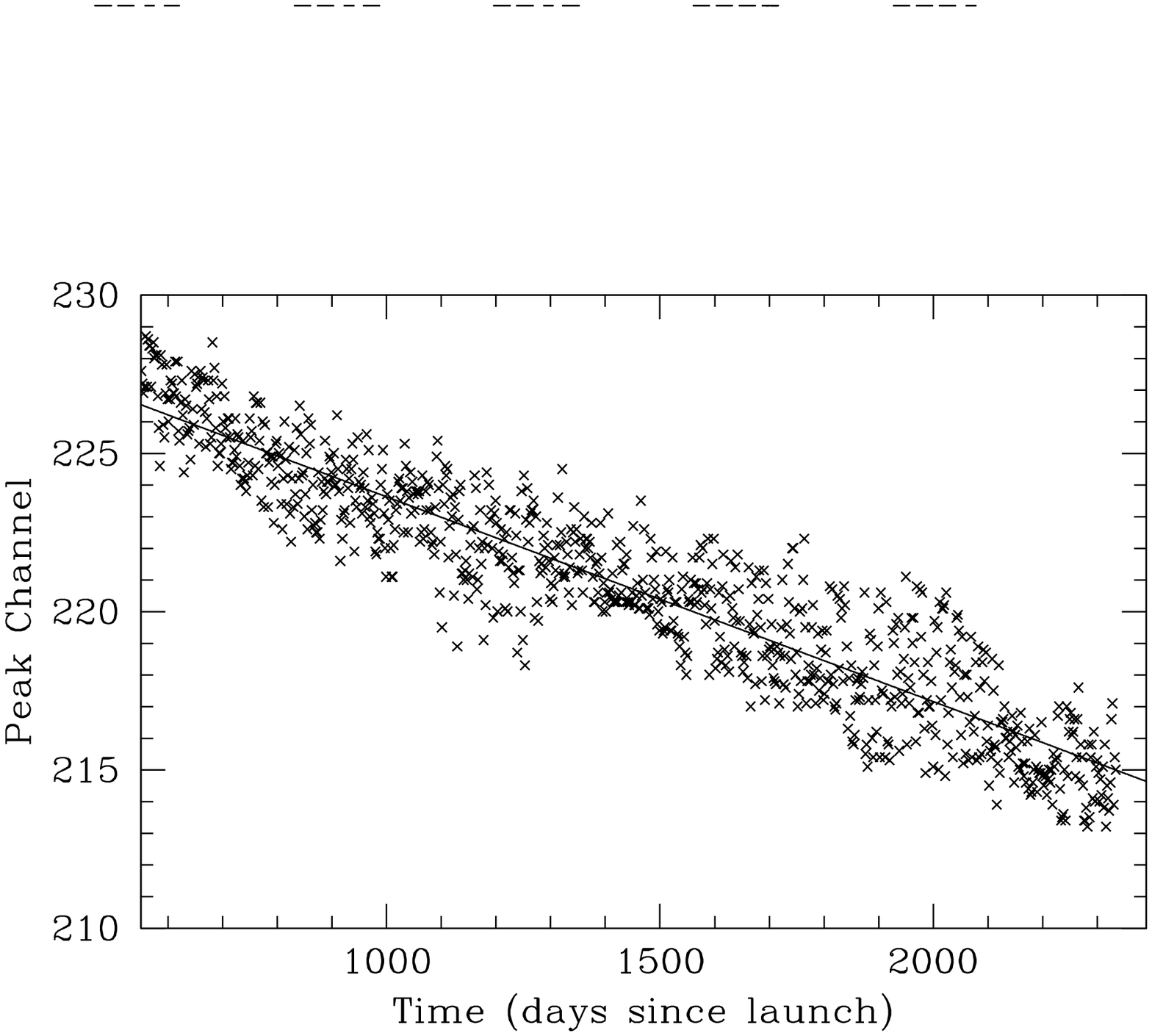}
\caption{The position of 30 keV peak in the calibration source as a function
of time since 20 March 2017, just after the last adjustment of HV.
The solid line shows a straight line fit to the data.}
\label{fig:gain}
\end{figure}

Another improvement is made in applying the gain shift to the background
spectrum to match that in the source spectrum. This was done using the measurement
of gain obtained from the calibration source in veto anode A8. It was found
that apart from a steady linear trend the gain also has some fluctuations
of the order of 2 channels (out of 1024). If the calculated gain shift is not correct
it can introduce some features around 30 keV where there is a prominent peak
in the background spectrum due to Xe K florescence. To improve the situation
for LAXPC20 data taken after  1 April 2017 (after the last HV adjustment
in the detector on 16 March 2017) the gain is fitted to a straight line and gain shift
between the background and source spectrum is calculated using this fit.
Figure~\ref{fig:gain} shows the peak position of 30 keV line in LAXPC20
after the last adjustment and the straight line fit to it. The fit gives
a variation of 1 channel in the peak position in about 150 days. Since typical
separation between the background and source observations is less than one month, it
gives a shift of less than 0.2 channels in most cases, which should not introduce any
artifact. This measure is not applied to older data and to other two
detectors as there the slope of gain change was much larger and adjustment
in HV or gas purification were more frequent \citep[figure 2 of][]{antia21}.



\section{Summary}

The LAXPC background shows a prominent quasi-diurnal variation which was not
accounted for in the earlier background models. Using over 5 years of data
the period of the quasi-diurnal variation is determined to be 84495 s,
which is the diurnal variation corrected for the precession of the orbit.
Identification of this period also made it possible to identify the orbits
where a large deviation was observed just after exit from SAA.
After incorporating this period in the background fit and removing some
interval from GTI, there is a significant
improvement in the background model and this model is incorporated in
the latest version of the software.

\acknowledgments

We acknowledge the strong support from Indian Space Research Organization (ISRO) in various aspect of instrument building, testing, software development and mission operation and data dissemination.



\end{document}